\documentclass[fleqn,usenatbib]{mnras}

\usepackage{newtxtext,newtxmath}

\usepackage[T1]{fontenc}

\DeclareRobustCommand{\VAN}[3]{#2}
\let\VANthebibliography\thebibliography
\def\thebibliography{\DeclareRobustCommand{\VAN}[3]{##3}\VANthebibliography}

\usepackage{graphicx}	
\usepackage{amsmath}	

\title[Simulation based parameter space for shock]{Simulation based parameter space for shock in transonic, sub-Keplerian accretion flow onto non-rotating black holes}

\author[Dasadhikary \& Garain]{
Aishi Dasadhikary,$^{1}$\thanks{E-mail: ad21ms075@iiserkol.ac.in}
Sudip K Garain,$^{1,2}$\thanks{E-mail: sgarain@iiserkol.ac.in}
\\
$^{1}$Department of Physical Sciences, Indian Institute of Science Education and Research Kolkata, Mohanpur, Nadia, WB - 741246, India\\
$^{2}$Center of Excellence in Space Sciences India, Indian Institute of Science Education and Research Kolkata, Mohanpur, Nadia, WB - 741246, India}

\date{Accepted XXX. Received YYY; in original form ZZZ}

\pubyear{2026}

\begin{document}
\label{firstpage}
\pagerange{\pageref{firstpage}--\pageref{lastpage}}
\maketitle

\begin{abstract}
Non-dissipative, transonic, sub-Keplerian accretion flow onto black holes is
characterized by two conserved parameters: specific energy and
specific angular momentum of the flow. For certain range of
these parameters, the accretion flow shows shock formation
and the post-shock matter forms a boundary layer which is believed to
shape the radiative properties of the accretion disk. In this work,
we identify the parameter space for shock in such accretion flows
using multi-dimensional numerical simulations around non-rotating black
holes and demonstrate that
the shock formation parameter space is much larger than the
analytically calculated one. We also find the boundary layer to
be dynamic for a significant part of this parameter space and
self-consistently produce outflow from the accretion disk.
\end{abstract}




\section{Introduction} 
\label{sec:1}

Black hole accretion solutions are necessarily transonic.
It has been shown that general solutions of non-dissipative
transonic accretion flows are described by two conserved
parameters, namely, the specific (per unit mass) energy $\epsilon$ and 
the specific angular momentum $l$ \citep{fukue1987,Chakrabarti1989a,
Chakrabarti1990b}. The solution with non-zero
$\epsilon$ but zero $l$ leads to the well known spherically
symmetric, Bondi accretion solution \citep{bondi1952}. 
With increasing $l$, the axisymmetric accretion flow
experiences increasing outward centrifugal force. This
results in the formation of a potential barrier
close to the black hole, very much similar to what is observed
in studying the nonradial motion of a test particle around 
black holes \citep{gravi,st1983}.
If the flow has sufficient energy to overcome the
barrier, accretion onto black holes is possible.
Because of this potential barrier,
accreting matter does not achieve a monotonically increasing
magnitude of radial velocity profile. Rather, the velocity 
profile shows a dip close to the black hole. For certain ranges
of $l$ and $\epsilon$, the accretion flow can even develop a shock.

Since the conceptualization of the shock formation in the
axisymmetric accretion flow, several analytical and numerical studies
have been carried out to investigate its formation, stability,
flow parameter dependence, etc. An analytical solution aims to
find the radial profile of accretion flow variables (e.g., velocity,
density, pressure, etc.) under the assumption of a thin,
conical or constant height flow
geometry, or for a thick flow under vertical equilibrium
\citep{Chakrabarti1989a,Chakrabarti1990b,Chakrabarti1993a,
Chakrabarti2001a}. The shock
location is found by Rankine-Hugoniot (RH) analysis, suitably performed
for these model flows. In numerical simulation, the solution
of time-dependent fluid dynamic equations using suitable
matter inflow boundary conditions determines the flow profile.
Inflowing fluid variables may be computed using the above-mentioned
analytical methods. For suitable parameters, the numerical solution
self-consistently identifies the shock solution. Comparison between
model-based analytical and numerical solutions shows very good
agreement between the two methods \citep{Chakrabarti1993a,Molteni1996b,
Lee2011a,garain2023,debnath2024}. Several general relativistic
hydrodynamics simulations \citep{kgbc2017,kgcb2019,garain2025} 
also demonstrate such agreement. Numerical simulations, 
additionally, allow us to study the long-time stability of the
shock solutions in multiple dimensions.

Shock formation in the transonic, axisymmetric accretion
disk plays an important role in interpreting observed high-energy radiation
from several black hole X-ray binaries. The sub-Keplerian component
in a two-component advective flow \citep[TCAF,][]{Chakrabarti1995a}
undergoes a shock transition. The shock makes the post-shock flow
geometrically thick and hotter, and allows efficient mixing
of the Keplerian component with the sub-Keplerian one to make
this region optically slim. Thus, this post-shock flow
becomes an ideal site for intercepting a fraction of low-energy photons
originating from the Keplerian disk and inverse-Comptonizing them
to higher energies. The shock-compressed post-shock flow
also allows the formation of outflow. TCAF model, as included
in XSPEC \citep{Arnaud1996a}, is being extensively
used to model X-ray observations from high-energy sources
\citep{Debnath2014a,Chakrabarti2015a,molla2017,shang2019,nandi2021,nath2024,eze2025}.

Shock formation in the transonic accretion solution requires the
presence of more than one X-type sonic point
\citep{fukue1987,Chakrabarti1989a}. 
Sub-sonic flow from an infinite distance first accelerates to super-sonic
speed by crossing the outer X-type sonic point. This super-sonic flow
then undergoes a shock transition, and the post-shock flow
again accelerates to super-sonic speed (i.e., passes through the inner
X-type sonic point) before reaching black hole
horizon (satisfying inner boundary condition).
Parameter space spanned by $\epsilon$ and $l$ shows
that such a condition of more than one X-type sonic point 
is satisfied for a limited range of these parameters
(see, e.g., \citet{Chakrabarti1990b}, \citet{das2001}).
Even within this limited region, the occurrence of shock in the 
accretion solution requires that the entropy of flow at the outer 
sonic point be less than that at the inner sonic point. 
This imposes further restrictions and makes the parameter space 
for the shock even more shrunk \citep{das2001}.
Moreover, RH analysis can locate a shock
that is standing (i.e., stationary). On the other hand, 
multi-dimensional numerical simulations indicate that a shock can 
form in non-dissipative flow
with parameters outside this allowed parameter space 
\citep{Molteni1994a,rcm1997,molteni2006,Giri2010a,garain2023}.
Not only this, the shock surface
can oscillate both radially as well as vertically 
\citep{deb2016,garain2023}. There are also instances of multiple
shock formation inside the accretion solution.
Such oscillating shocks are believed to produce quasi-periodic
oscillations (QPOs) in the TCAF model 
\citep[][and references therein]{Chakrabarti2015a}.
However, such simulations are carried out for a few sets of
parameters outside the said parameter space in isolated ways.

Several other
groups independently investigated the non-dissipative sub-Keplerian accretion
solutions using numerical simulations and found a variety of
solutions. General relativistic (GR) hydrodynamic simulations of
\citet{wilson1972,hawley1984} find shocked accretion solutions,
though the shocks are unstable and move outward. 
\citet{ryu1995apj} considers accretion onto a Newtonian star using
a non-GR code and reports standing as well as oscillating
accretion shock solution. \citet{proga2003hd, proga2003mhd} study
the accretion of low-angular momentum flow onto a non-rotating
black hole described by pseudo-Newtonian potential and find formation
of a non-accreting torus close to the equator. 
In these simulations, infalling matter starts accreting from the outer
boundary with a small, latitude-dependent $l$ with $l$ 
being highest at the equator and zero near the pole.
A similar setup is used in GR hydrodynamics
simulations of \citet{sukova2017} and 
\citet{olivares2023}. These simulations show a variety of solutions
with various choices of $l$, and find shock and post-shock
torus formation for some cases. There are non-GR and 
GR magneto-hydrodynamics
simulations also studying the flow profile, including shock of low angular
momentum accretion onto black holes \citep{mhd2020,mao2025}

All the above-mentioned simulation works thus agree that for a certain
range of flow parameters, shocks can form in the accretion flow.
Model-based analytical works have identified the parameter space for
standing shock formation. However, numerical simulations seem to 
indicate that various types of shock solutions (standing, propagating,
oscillating, etc.) are possible. In this work, we perform a 
thorough investigation using 
multi-dimensional numerical simulation on a much larger parameter
space than the theoretically predicted space
and aim to identify the parameter space which shows shock formation
in the accretion solution. To our knowledge, no such
simulation-based parameter space identification for shock has been done
earlier. We use the analytically calculated parameter space
as our guide and perform our scan inside a rectangular region 
surrounding this. We choose more than 280 sets of ($l,\epsilon$)
pairs inside this rectangular
region and perform a two-dimensional simulation (assuming axisymmetry)
of a geometrically thick accretion disk corresponding to each set.
We perform systematic analysis on this simulated accretion disk
configurations to identify whether it is accretion or outflow
dominated, and if it is accretion dominated, how does the
shock surface behave (in case shock forms).
Based on our findings, we mark a set, and finally, we draw
the boundary of the parameter space for the shock in the accretion
flow. We also mark the no-shock, oscillatory shock
and outflow-dominated regions.

Our paper is organized as follows: In Section~\ref{sec:2}, we review
the analytical method for identifying the ($l,\epsilon$) parameter
space for shock formation in the transonic, sub-Keplerian
accretion flow. In Section~\ref{sec:3}, we present our multi-dimensional
simulation procedure. In Section~\ref{sec:4}, we present our results.
Finally, in Section~\ref{sec:5}, we provide a summary and our concluding remarks.

In our following calculations, we use $r_g=2GM_{bh}/c^2$ as unit
of distance, $r_g/c$ as unit of time and $r_gc$ as unit of 
specific (i.e., per unit mass) angular momentum. 
Specific energy is measured in units of $c^2$.
Here, $G$ is the gravitational constant, $M_{bh}$ is the
mass of the black hole and $c$ is the speed of light in vacuum.

\section{Analytical method for parameter space identification}
\label{sec:2}
In this section, we briefly describe the methods to
identify the parameter space for shock formation. 
Several works describe the parameter space calculation method
for non-dissipative transonic accretion flow onto non-rotating
black holes \citep{Chakrabarti1989a, Molteni1994a, das2001, 
mondal2006, mondal2020}. We use a
semi-analytic method for reproducing the parameter space.

Non-dissipative flow is described by the following energy
and mass conservation equations:
\begin{equation}
    \epsilon = \frac{v_e^2}{2} + \frac{a_e^2}{\gamma - 1} + \frac{l^2}{2r^2} -\Phi
    \label{eq:enrg_cons}
\end{equation}
\begin{equation}
    \dot M = v_e \rho r h
    \label{eq:mass_cons}
\end{equation}

Here, $r$ is the distance from the black hole, $v_e(r)$ is the 
fluid velocity on the equator, $a_e(r)$ is the sound speed at 
the equator, $\rho(r)$ is the density and $h(r)$ is the disk half-thickness.
$\Phi(r)$ represents the gravitational potential.
For current work, we use $\Phi(r)=-1/2(r-1)$ \citep{Paczynsky1980a}.
We perform the calculation for the vertical equilibrium model and
hence, use $h(r) = a_e r^{\frac{1}{2}}(r - 1)$ \citep{Chakrabarti1990b}.
These equations are used along with the equation of state
\begin{equation}
    P = K \rho^\gamma,
\end{equation} \\
with $P$ being the pressure, $K$ being a constant measuring
the entropy and $\gamma=4/3$ being the polytropic index.
This equation of state provides us with 
\begin{equation}
    a_e^2 = \frac{\gamma P}{\rho} = \gamma K \rho^{(\gamma-1)}
\end{equation}

This equation allows to substitute $\rho$ form Equation \ref{eq:mass_cons}
and express $\dot M$ in terms on $v_e$ and $a_e$. Next, this equation
and Eq. \ref{eq:enrg_cons} are differentiated w.r.t $r$ keeping in
mind that both $d\epsilon/dr=0$ and $d\dot M/dr=0$. Now,
eliminating $d a_e/d r$ from both equations, one can obtain
\begin{equation}
	\frac{dv_e}{dr} = \frac{\frac{2a_e^2}{\gamma + 1}\cdot\frac{5r - 3}{2r(r-1)} - {\frac{1}{2(r - 1)^2} + \frac{l^2}{r^3}}}{v_e - \frac{2a_e^2}{(\gamma + 1)v_e}}
    \label{eq:dvdr}
\end{equation}

\begin{figure}
\centering
\includegraphics[width=\columnwidth]{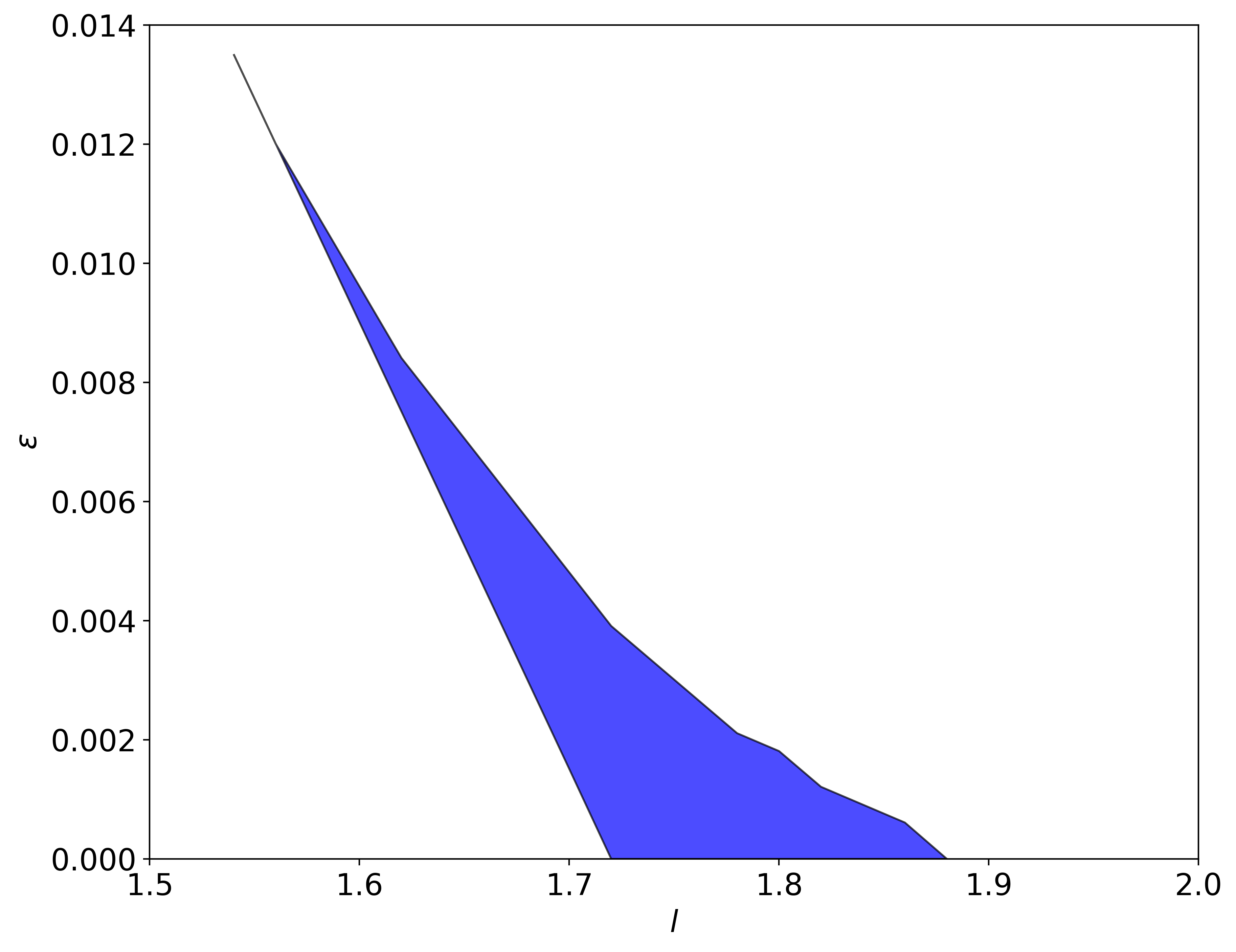}
\caption{ shows allowed parameter space in the $l-\epsilon$ plane for
the standing shock formation in the non-dissipative sub-Keplerian
accretion flow. This is found using analytical method.}
\label{fig1}
\end{figure}

One can now perform the so-called sonic (critical) point analysis
on this equation.
This yields 
\begin{equation}
v_{e,c}^2=\frac{2}{(\gamma + 1)}a_{e,c}^2
    \label{eq:sonic1}
\end{equation}
and
\begin{equation}
	a_{e,c}^2=\frac{(\gamma + 1)(r_c - 1)}{r_c^2(5r_c - 3)}\left[\frac{r_c^3}{2(r_c-1)^2} - l^2\right]
    \label{eq:sonic2}
\end{equation}
at the critical point $r=r_c$. Thus, at the critical point we have
\begin{equation}
	\epsilon = \frac{v_{e,c}^2}{2} + \frac{a_{e,c}^2}{\gamma - 1} + \frac{l^2}{2r_c^2} -\frac{1}{2(r_c-1)}
\end{equation}
Upon using Eq. \ref{eq:sonic1} and Eq. \ref{eq:sonic2} in this equation, one obtains a quartic equation in $r_c$. 
The solution allows for locating the critical points: for a given
set of $l$ and $\epsilon$, one can solve this equation analytically
or numerically, and identify
whether the given ($l, \epsilon$) pair allows the existence of three
critical points outside the black hole horizon. Out of these three
critical points, the outermost and innermost points are X-type,
and the middle one is O-type. If a pair
satisfies this condition, we proceed to compute the solution and
identify the shock location. Eq. \ref{eq:dvdr} can now be solved
numerically using fourth-order Runge-Kutta scheme
to obtain $v_e(r)$ and subsequently, $a_e(r)$ and hence the Mach
number $M(r)=v_e/a_e$. Solution branches passing through both 
the X-type critical points are found. Next, we use the shock
invariant combination of Mach numbers (following \citet{Chakrabarti1989a}),
which is continuous across the shock, to find the existence and
location of the shock. To identify the shock formation
parameter space spanned by ($l, \epsilon$), we scan a
rectangular domain [1.4:2.0]$\times$[0.00001:0.0177] using
31$\times$60 pairs and identify the pairs that allow shock formation.
The shaded region in Fig. \ref{fig1} shows the said parameter space.

Stability analysis of accretion flows involving shocks 
has been performed by several groups. \citet{Chakrabarti1989a} has
performed a local stability analysis and demonstrated that shocks
are stable in the short-wavelength limit. Several other subsequent
global stability analysis \citep{Nakayama1992a, Nakayama1994a, 
Nobuta1994a, Gu2003a, Gu2006a} demonstrate that these shock structure
may be stable under various conditions. Two-dimensional, time-dependent
numerical simulations of shocked accretion flows in the equatorial 
plane with nonaxisymmetric azimuthal perturbations reveal that axisymmetric
shock structure becomes nonaxisymmetric \citep{Molteni1999a}. However, 
this nonaxisymmetric, spiral shock structure is stable. Such simulations
are extended to three-dimensions in \citet{garain2023} where we again
confirm the same conclusion.

\section{Simulation procedure}
\label{sec:3}

For this study, we run more than 280 number of two-dimensional 
ideal hydrodynamics
simulations in the $R-Z$ cylindrical domain. We assume axi-symmetry in the
$\phi$ direction. The following conservation equations
are solved numerically:

\begin{equation}
\frac{\partial \mathbf{U}}{\partial t} + \frac{1}{R}\frac{\partial \left(R\mathbf{F_R}\right)}{\partial R} + \frac{\partial \mathbf{F_Z}}{\partial Z} = \mathbf{S},
\label{eq1}
\end{equation}
where,
the vector of conserved variables $\mathbf{U}$, R-flux $\mathbf{F_R}$,
and Z-flux $\mathbf{F_Z}$ can be written as:
\begin{eqnarray*}
{\bf U }= \left(
\begin{array}{ccccc}
\rho \\ \rho v_R \\ \rho l \\ \rho v_Z \\ E 
\end{array}
\right);
\quad
{\bf F_R}=\left(
\begin{array}{ccccc}
\rho v_R \\ \rho v_R^2 + P \\ \rho l v_R \\ \rho v_R v_Z\\ \left( E + P\right)v_R
\end{array}
\right);
\quad
{\bf F_Z}=\left(
\begin{array}{ccccc}
\rho v_Z \\ \rho v_R v_Z \\ \rho l v_Z \\ \rho v_Z^2 + P \\ \left(E + P\right)v_Z
\end{array}
\right)
\end{eqnarray*}
And the source term is:
\begin{eqnarray*}
{\bf S }&=& \left(
\begin{array}{ccccc}
0 \\ \frac{\rho v_\phi^2}{R} + \frac{P}{R}- \rho \frac{\partial \Phi}{\partial r}\frac{R}{r} \\ 0 \\
- \rho \frac{\partial \Phi}{\partial r}\frac{Z}{r} \\
- \rho \frac{\partial \Phi}{\partial r}\frac{\left(R v_R + Z v_Z\right)}{r}
\end{array}
\right).
\end{eqnarray*}
Here, $\rho$ is density, $v_R, v_\phi, v_Z$ are three components of
velocity, $P$ is pressure, $E=\frac{1}{2}\rho (v_R^2 + v_\phi^2 + v_Z^2)
+ \frac{P}{\gamma - 1}$ and $l=Rv_\phi$. $r$ represents the spherical
radius and is given by $r=\sqrt{R^2 + Z^2}$. $\Phi(r)=-1/2(r-1)$
represents the gravitational potential. We use polytropic
equation of state $P=K \rho^\gamma$ with $\gamma=4/3$ and $K$ being
a measure of entropy.
The vector of primitive variables is denoted by 
\begin{eqnarray*}
{\bf V }&=& \left(
\begin{array}{ccccc}
\rho \\ v_R \\ v_\phi \\ v_Z \\ P 
\end{array}
\right).
\end{eqnarray*}

\begin{figure}
\centering
\includegraphics[width=\columnwidth]{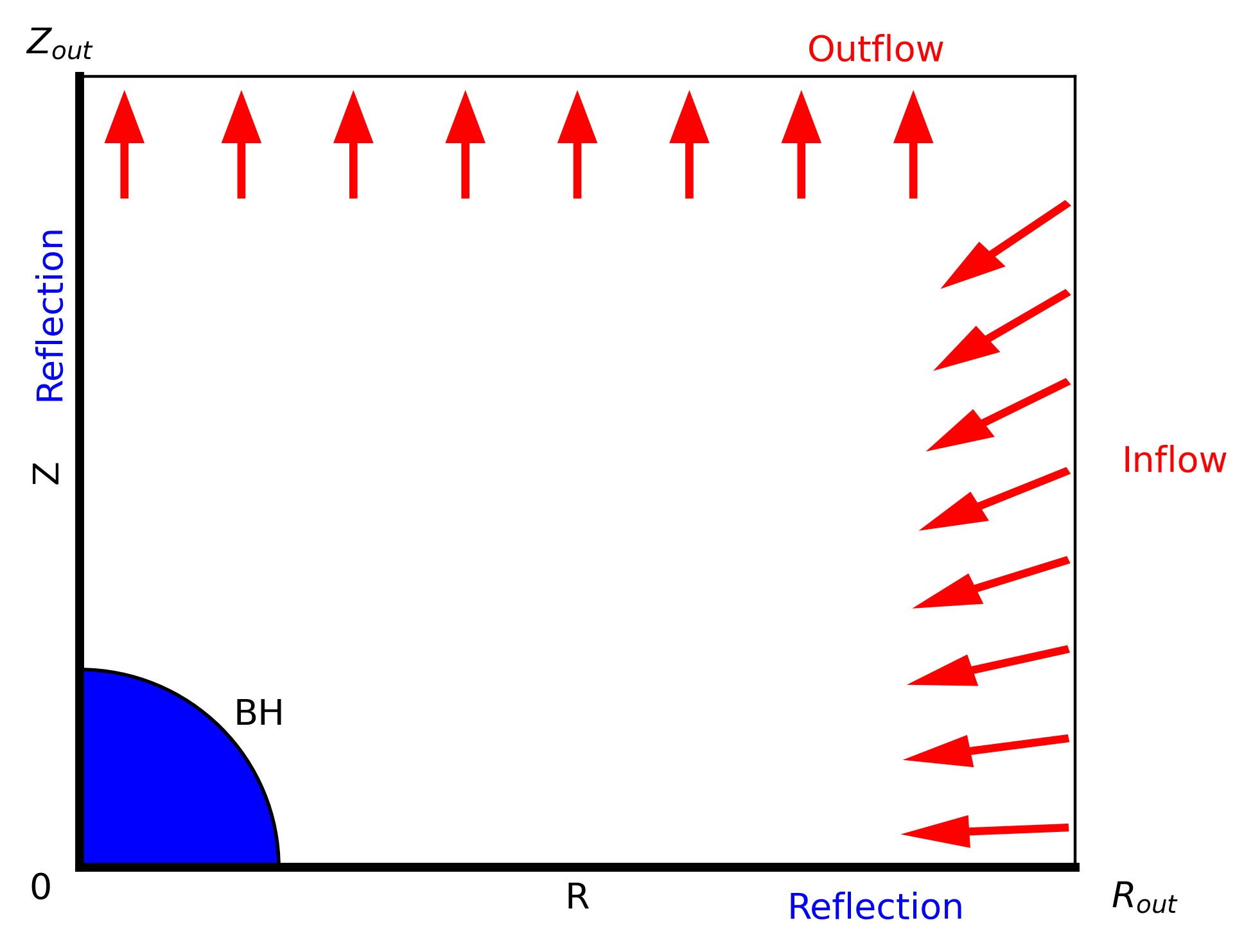}
\caption{ shows the simulation domain and the set up. Inflow
boundary condition is used at $R=R_{out}$ and outflow boundary
condition is used at $Z=Z_{out}$. Matter is absorbed inside 
$r=r_{BH}$ (blue region). Reflection boundary conditions are
used on the axis (R=0) and the equator (Z=0).
}
\label{fig:simulation_frame}
\end{figure}

We use similar simulation set up which has been used over decades to simulate
such transonic accretion flow \citep{cm1993,ryu1995apj,Molteni1996b,
Chakrabarti2004a,otm2007,Giri2010a,om2012,Garain2012a,Das2014a,ggc2014,
Lee2016a,kgbc2017,kgcb2019,garain2023,debnath2024}.
Our simulation set-up is sketched in Fig. \ref{fig:simulation_frame}.
The simulation domain is [$0:R_{out}$]x[$0:Z_{out}$].
A black hole with radius $r_{BH}=\sqrt{R^2+Z^2}=1.8$ is placed at
the origin. To mimic matter absorption by the black hole,
we set $\rho$ and $P$ to respective floor values and the velocity
components to zero on all grid points inside $r=r_{BH}$.
We allow matter coming from far away to enter our simulation domain
through the outer radial boundary. Thus,
we impose an inflow boundary condition at $R=R_{out}$, i.e.,
we maintain a constant {\bf V} on the ghost zones of
this boundary. The disk
half-height at injection is assumed to be $h=Z_{out}=R_{out}/2$.
Reflection boundary condition is used on the axis $R=0$
and on the equator $Z=0$. Because our flow is axisymmetric,
the velocity components normal to the rotation axis cancel out.
This justifies reflection boundary condition
on the axis. For our flow, matter rarely reaches $R=0$
because of non-zero angular momentum. Hence, the flow profile
is generally insensitive to this specific boundary condition, 
which is verified for a few cases. Reflection boundary condition
on the equator $Z=0$ arises due the assumption that
the flow dynamics is symmetric
with respect to the equator. Thus, the velocity components normal 
to the equator is exactly zero at $Z=0$ leading to zero
mass flux across the equator. This assumption imposes some restriction
though as discussed in Section \ref{sec:5}. 
The outflow boundary condition is used
on the upper Z-boundary $Z=Z_{out}$. The simulation domain is
initially filled with static background matter having
density and pressure as respective floor values.

Our numerical solution scheme is the same as in \citet{garain2023}.
We use the finite volume method, and the code is globally second-order accurate. The spatial reconstruction
is done using the van Leer limiter, interfacial fluxes are calculated
using the HLL Riemann solver, and the time-integration is done using
a two-stage strong stability preserving Runge-Kutta scheme.

Since the numerical solutions depend on the inflow
boundary condition, we describe its implementation in details
here. The inflow parameters {\bf V} need to be found at all radial
and vertical ghost zones adjacent to the outer radial boundary
at $R=R_{out}$. In the radial direction, we use two ghost zones,
and we use the same {\bf V} in both. However, {\bf V} varies in the
vertical direction. We first compute {\bf V} at the equator.
$v_e$ and $a_e$ are computed following Section \ref{sec:2}
for a given ($l,\epsilon$) pair. We set $v_R=v_e$ and $v_Z=0$.
Since the simulations are non-dissipative, density $\rho=\rho_e$
is scaled to 1. This allows us to compute pressure
$P=P_e=a_e^2 \rho_e/\gamma$. $l$ value allows to calculate 
$v_{\phi}=l/R_{out}$.
Once {\bf V} at the equator is set, we proceed to compute
its values at other heights. Note that if we use the equatorial
{\bf V} at different heights, resulting $\epsilon$ will not 
remain constant due to reducing gravitational energy $\frac{1}{2(r-1)}$
with heights. To compensate for this, we reduce
$v_r=\sqrt(v_R^2 + v_Z^2)$ and $a$ with increasing height
such that the $v_r/a$ ratio remains the
same as the equatorial Mach number $v_e/a_e$. This ensures that
the injected flow has a constant $\epsilon$. $v_R$ and $v_Z$
are chosen in a way that $v_r$ vector points to the central black
hole. Next, we enforce the flow to be isentropic at all heights.
This allows us to find the density $\rho$ at all heights using
$\rho=\left(\frac{a^2}{\gamma K_e}\right)^{1/(\gamma -1)}$,
with $K_e$ being the equatorial entropy $K_e=P_e/\rho_e^\gamma$.
Using this $\rho$, $P=K_e \rho^\gamma$ is found at all heights.
$v_{\phi}$ is computed using $v_{\phi}=l/R_{out}$.

\section{Results}
\label{sec:4}

Using the above numerical simulation scheme and boundary conditions,
we aim to find a parameter space analogous to the one shown
in Fig. \ref{fig1}. For this,
we scan the ($l,\epsilon$) domain [1.4:1.9]x[0.0001:0.0061]
using judiciously chosen 286
points (each point corresponds to a ($l,\epsilon$) pair).
The outermost X-type critical radius ($r_c$) is calculated for
a given point, and it is used to determine the simulation domain. 
$R_{out}$ is chosen to be just inside this $r_c$ and $Z_{out}=R_{out}/2$.
Thus, we don't simulate a flow passing through the outer sonic point.
Rather, matter starts with a supersonic speed from the
outer boundary. This supersonic injection ensures that the inflow
properties remain unaffected from the acoustic feedback from 
the interior of the simulation domain and correctly represent the 
flow solution corresponding to given conserved inflow parameters.
This practice is followed in all the references mentioned above.
Clearly, our simulation domains vary with the points. We adjust
number of grids in the $R$ and the $Z$ directions accordingly.
For the chosen range of parameters, the outermost $r_c$ ranges
between 34.4 and 2995.4. Thus, the lowest value of $R_{out}$ across
our simulations is 29. However, due to limited computational
resources, the highest value of $R_{out}$ across our simulations 
is restricted to 250.
For all simulations, we use ratioed grids in the $R$ and $Z$
with a common ratio of 1.003.
Before fixing the grid numbers, we experimented with a few cases
and confirmed that the final solution doesn't depend on the grid size
(i.e., solution is converged.)
All simulations have been run till $t=50000$.

\subsection{Parameter space}

\begin{figure*}
\centering
\includegraphics[width=\textwidth]{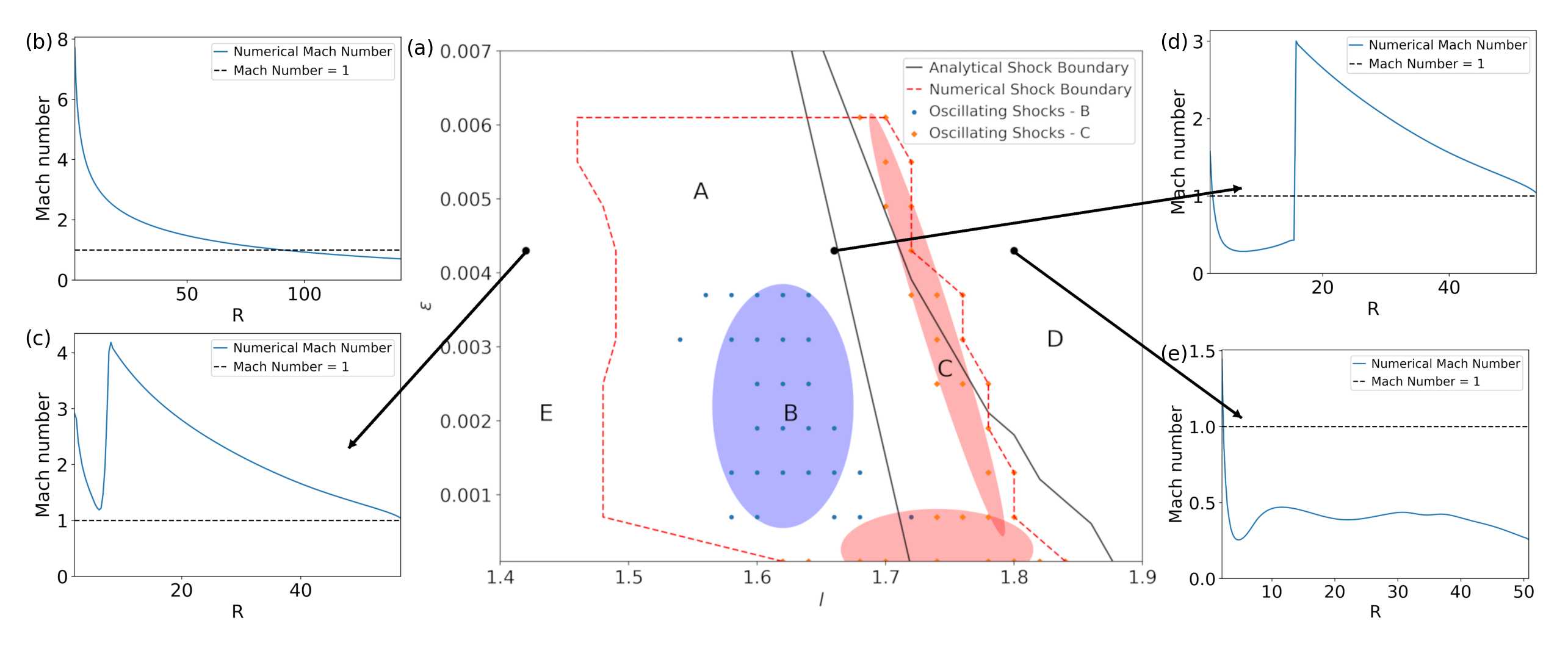}
\caption{Classification of parameter space for the non-dissipative
sub-Keplerian accretion flow using numerical simulation. The red dashed
line in (a) surrounds the parameters which show shock formation. 
The solid black lines overplot the boundary of the shaded region shown
in Fig. \ref{fig1}. Shocks can be standing or oscillatory. 
Region marked \textbf{A} shows standing shock solutions. Numerical 
simulations identify different types of oscillatory shocks 
in regions marked \textbf{B} (blue) and \textbf{C} (pink).
Parameters from region \textbf{E} do not show shock formation
though the effect of centrifugal barrier may be present. Accretion 
flow with parameters from region \textbf{D} passes through inner 
sonic point only.
Four plots surrounding (a) demonstrate the gradual change in the
solution nature with increasing $l$ for a given $\epsilon=0.0043$
value. Radial Mach number
variation along the equator is plotted in these plots. (b) shows
the standard Bondi solution profile for $l=0$. With increasing $l$,
the centrifugal barrier becomes prominent. (c) shows the solution
with the barrier but without a shock for $l=1.42$. (d) shows the solution
with a shock for $l=1.66$. Finally, with much higher $l=1.8$, the solution
shows presence of only the inner sonic point.
	}
\label{fig:overall_picture}
\end{figure*}

We run the multi-dimensional simulation for each point of
the parameter space and confirm that it reaches a steady state
by simulation stopping time.
During the run, we compute the shock location on the equatorial
plane $Z=0$. This time-series data allows us to identify whether
a shock is formed for this point, and if yes, whether it is standing
or time varying. This way, we categorize all 286 points based on
the solution characteristics and identify a bulk region on
the ($l,\epsilon$) plane that shows shock formation.
Area inside the red-dashed curve in Fig. \ref{fig:overall_picture}(a)
represents this region. The analytically determined parameter space
for the standing shock is also shown here by the black solid line.
Note that, unlike the smooth boundary found using the analytical method,
the numerically identified boundary is slightly jagged.
This is the manifestation of scanning the ($l,\epsilon$) domain 
with finite number of points. One may find smoother boundary by 
increasing the number of scanning points.

We identify this shock parameter space boundary curve
at the lower and upper $l$ values by analyzing the gradual change
of the solution nature. Four plots surrounding 
Fig. \ref{fig:overall_picture}(a) demonstrates this gradual 
change in the solution nature because of increasing $l$ for a 
given $\epsilon=0.0043$ value. Radial Mach number
variation along the equator is plotted in these plots.
For a given value of $\epsilon$, the solution
corresponding to the lowermost $l$ value (i.e., $l=0$) is the standard
Bondi accretion solution having only one sonic point : a subsonic
flow far away from the black hole becomes supersonic after crossing
this sonic point. Fig. \ref{fig:overall_picture}(b) shows such solution for 
$\epsilon=0.0043, l=0$. For points just outside the left red-dashed 
line boundary,
the equatorial Mach number profiles show clear evidence of slow down due 
to centrifugal barrier, but the flow remains supersonic. This ensures
absence of inner sonic point for such parameters. Fig. 
\ref{fig:overall_picture}(c) shows such a solution for
$\epsilon=0.0043, l=1.42$. This region is marked as \textbf{E}.
For slightly higher $l$, the equatorial Mach profiles show steady 
supersonic to subsonic transition (shock solution) and subsequently
accreting flows pass through the inner sonic point before getting
accreted supersonically. Fig. \ref{fig:overall_picture}(d) shows 
such a solution for $\epsilon=0.0043, l=1.66$. For some cases,
the shock solutions are oscillatory. Region \textbf{A} shows the points
with steady shock solutions, whereas, regions \textbf{B} and \textbf{C}
show the oscillatory shock solutions. With increasing $l$ (larger 
centrifugal barrier), the shock moves away from the black 
hole. And finally, we find a situation when the shock reaches the 
outer boundary of the domain leaving behind the subsonic flow which
passes through the inner sonic point before getting accreted supersonically.
Fig. \ref{fig:overall_picture}(e) shows such a solution for
$\epsilon=0.0043, l=1.8$.
Points from region \textbf{D} (right to the upper $l$ boundary) show
these solutions. We discuss examples of solutions
from each of these regions separately below.

We observe that the simulation-based parameter space for shock 
is larger, specifically for lower $l$, than the analytical 
calculation-based parameter space. In analytical
calculations, the flow is assumed to be in
vertical equilibrium i.e., vertical pressure gradient force is balanced
by the vertical component of gravity. This balance does not allow vertical
motion of the fluid as the net force in the vertical direction is zero. 
In contrast, we find significant amount of vertical motion in our 
simulations, specifically just after the flow passes the centrifugal
barrier. This vertical motion accumulates a large part of the inflowing 
matter closer to the equatorial region, thereby enhance the pressure 
in the equatorial region, which helps in shock formation.
For example, a set of parameters, $\epsilon = 0.0055$
and $l=1.66$, allows a shock formation at $r=12.14$ as per analytical 
method and the density compression is found to be 2.11. However, the 
numerical simulations detects a steady shock at $R=16.06$ on the equator 
with a density compression of 3.7. This extra compression results in
extra pressure build up in the post-shock region. One may add this extra
pressure term parametrically in the Rankine-Hugoniot analysis to analytically
find the shock location as done in \citet{Molteni1994a} and
demonstrate that shock actually forms farther for a reasonable value
of the parameter. For this case, the parameter value is found to
be 0.1 which is reasonable (see \citet{Molteni1994a}). This same 
reason allows formation of shock in numerical simulations for the 
parameter sets with lower $l$ values, which do not accept shock solution 
analytically.

In passing, we also wish to mention that in full three-dimensional (3D)
simulation, flow is dynamically active in all three
spatial directions. However, if a full axisymmetric inflow
boundary condition is imposed, the flow remains axisymmetric throughout
the domain because the centrifugal barrier is axially symmetric
and fluid parcels along all $\phi$ are affected in the same
manner. This is demonstrated in \citet{garain2023}. Since the flow 
remains fully axisymmetric, there is no difference between our current 2D
and a similar 3D flow structure. So we believe that the shock
parameter space found in this work will remain the same if we repeat
the task using 3D simulation with full axisymmetric inflow
boundary condition. However, if non-axisymmetric variation (through 
e.g., a density perturbation) is explicitly
introduced in the $\phi$ direction of a 3D simulation, flow develops
 non-axisymmetric structure and solution profiles differ significantly
from a 2D flow structure. In several works \citep{Molteni1999a,
nagakura2008,nagakura2009,garain2023}, it is reported that
such a perturbed flow develop larger post-shock, turbulent region.
Thus, it is possible that the shock parameter space may change in 3D 
simulations with non-axisymmetric structure.

\subsubsection{Region \textbf{A}}
\begin{figure*}
\centering
\includegraphics[width=\textwidth]{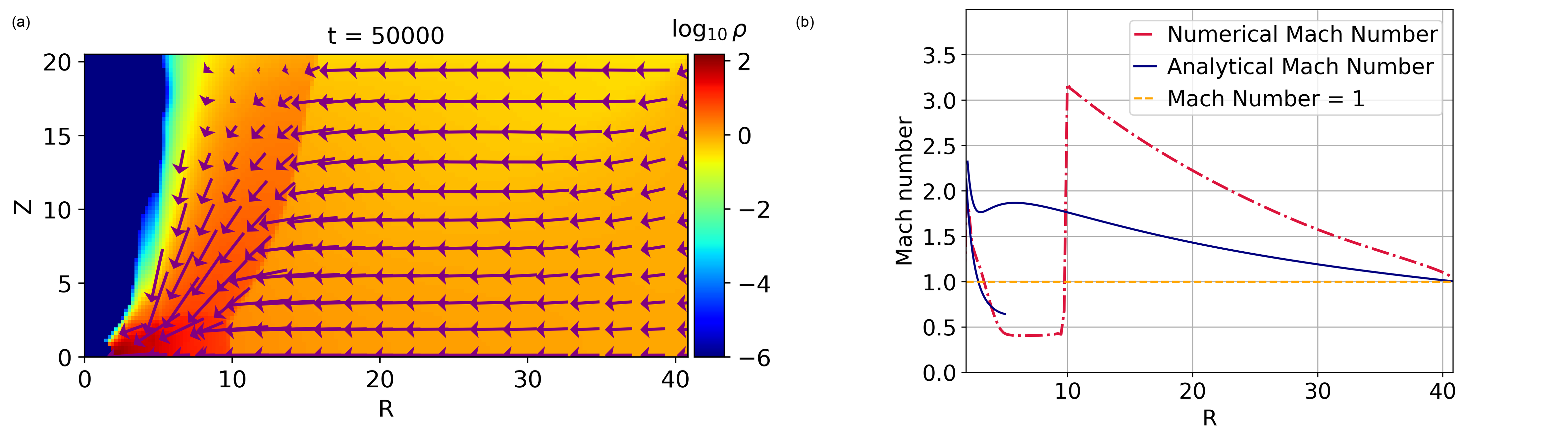}
\caption{(a)log$_{10}\rho$ distribution with velocity vectors overplotted
and (b) radial variation of Mach
number are shown here for a typical point of \textbf{A} region
in Fig. \ref{fig:overall_picture}(a). This is an example of standing
shock solution in our simulation. Parameters $\epsilon= 0.0055$
and $l=1.62$ are
used for this simulation. This set of parameters does not allow shock
formation as per analytical method. However, numerical simulation 
shows standing shock formation at $R=9.98$ on the equator.}
\label{fig:standing_shock}
\end{figure*}

Our numerical simulations find solutions with standing as well as
oscillatory shocks. For a significant part of the scanned space,
shocks are non-oscillatory or oscillatory with a very small
amplitude ($< 2 r_g$). We mark that region of space with \textbf{A}
in Fig. \ref{fig:overall_picture}(a). Typical solution profile
for a point in this region is shown in Fig. \ref{fig:standing_shock}.
$\epsilon= 0.0055$ and $l=1.62$ are used for this case.
This simulation has been conducted in a simulation domain
[0:40.82]$\times$[0:20.34] using [148 $\times$ 74] ratioed grids.
Fig. \ref{fig:standing_shock}(a) shows the log$_{10}$$\rho$ 
distribution in color, overplotted with velocity vectors.
The solution at time $t=50000$ is shown here. One can identify
the parabolic shock surface from the density contrast
starting at the equator
around $R=10$ and moving vertically up. This surface is known
as centrifugal barrier supported boundary layer (CENBOL).
Fig. \ref{fig:standing_shock}(b)
shows the radial variation of the Mach number along the equator
(dashed-dot line), and the shock location is $R=9.98$ on the equator.
We additionally plot the Mach number profile obtained from
the analytical method (solid line). Branches passing through the
outer and the inner sonic points are shown. However, the analytical solution
is shock free. The numerical solution shows shock solution and
post-shock flow follows the analytical branch passing through the
inner sonic point.
For all the other points in this region, the solution profile is
mostly similar except that the shock is located nearer or
farther from the black hole. We find that for a given $\epsilon$,
the shock location moves farther out with increasing $l$.
On the other hand, for a given $l$, the shock location moves
farther out with increasing $\epsilon$. 

\subsubsection{Region \textbf{E}}
\begin{figure*}
\centering
\includegraphics[width=\textwidth]{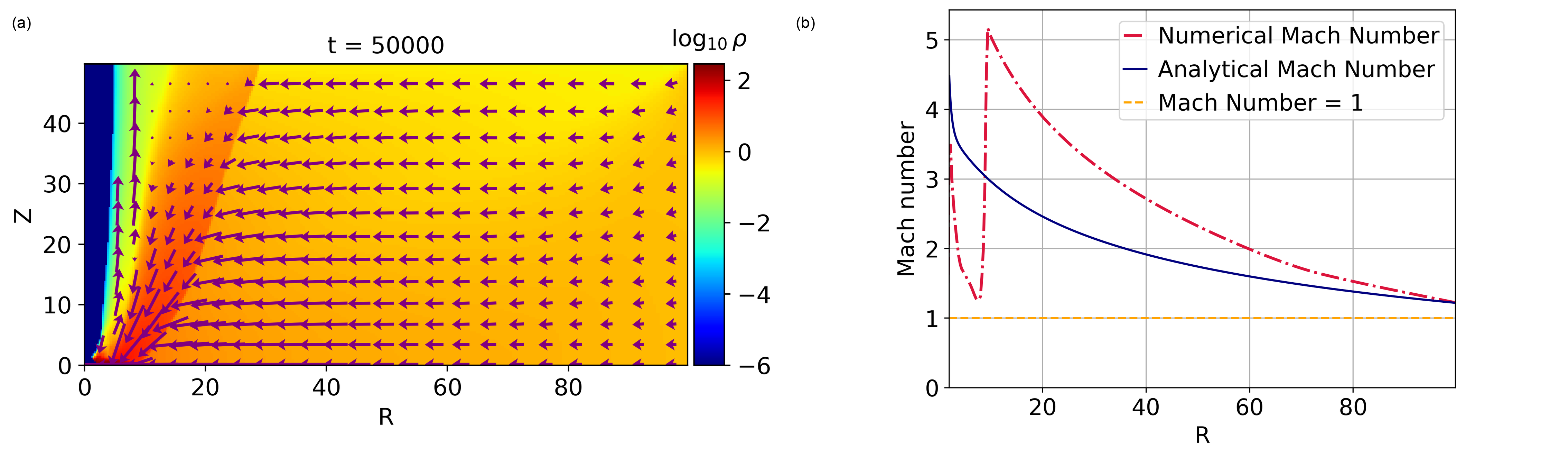}
\caption{(a)log$_{10}\rho$ distribution with velocity vectors overplotted
and (b) radial variation of Mach
number are shown here for a typical point of \textbf{E} region
in Fig. \ref{fig:overall_picture}(a). This is an example of no
shock solution. Parameters $\epsilon=0.0019$ and $l=1.46$ are
used for this simulation. This set of parameters does not show shock
formation both analytically and numerically. However, numerical simulation
shows piling up of matter and resulting supersonic to supersonic jump
due to centrifugal barrier.}
\label{fig:no_shock}
\end{figure*}

Solution profile for a typical point in region \textbf{E}
also shows the presence of CENBOL, albeit in a very weak form.
This is understandable because a lower $l$ would produce a weaker
centrifugal barrier. Fig \ref{fig:no_shock}(a) shows the log$_{10}$$\rho$
distribution in color, and (b) shows the Mach number profile 
along the equator (dashed-dot line)
for a typical point from this region. Here again, 
we show the Mach number profile obtained from
the analytical method (solid line). Only outer sonic point
exists for this case. $\epsilon=0.0019$ and $l=1.46$
are used for this case. 
The solution at time $t=50000$ is shown here.
This simulation has been conducted in a simulation domain
[0:99.72]$\times$[0:49.765] using [256$\times$128] ratioed grids.
We notice a very thin layer of enhanced
density behind CENBOL. The Mach number profile also shows a jump.
However, the flow remains supersonic even after that and accretes
onto the black hole supersonically. Thus, no shock (supersonic
to subsonic transition) forms for such cases and hence, the inner
sonic point is also not present. 

\subsubsection{Region \textbf{B} and \textbf{C}}
Parameters from regions marked \textbf{B} and \textbf{C} show
numerical solutions with the oscillatory shocks. In
Fig. \ref{fig:overall_picture}(a), we have marked only those
oscillating cases that have a significant amplitude, greater
than or equal to 2. However, the nature of the oscillations
is different for these two regions. 
We extensively discuss these solutions now.


\begin{figure}
\centering
\includegraphics[width=\columnwidth]{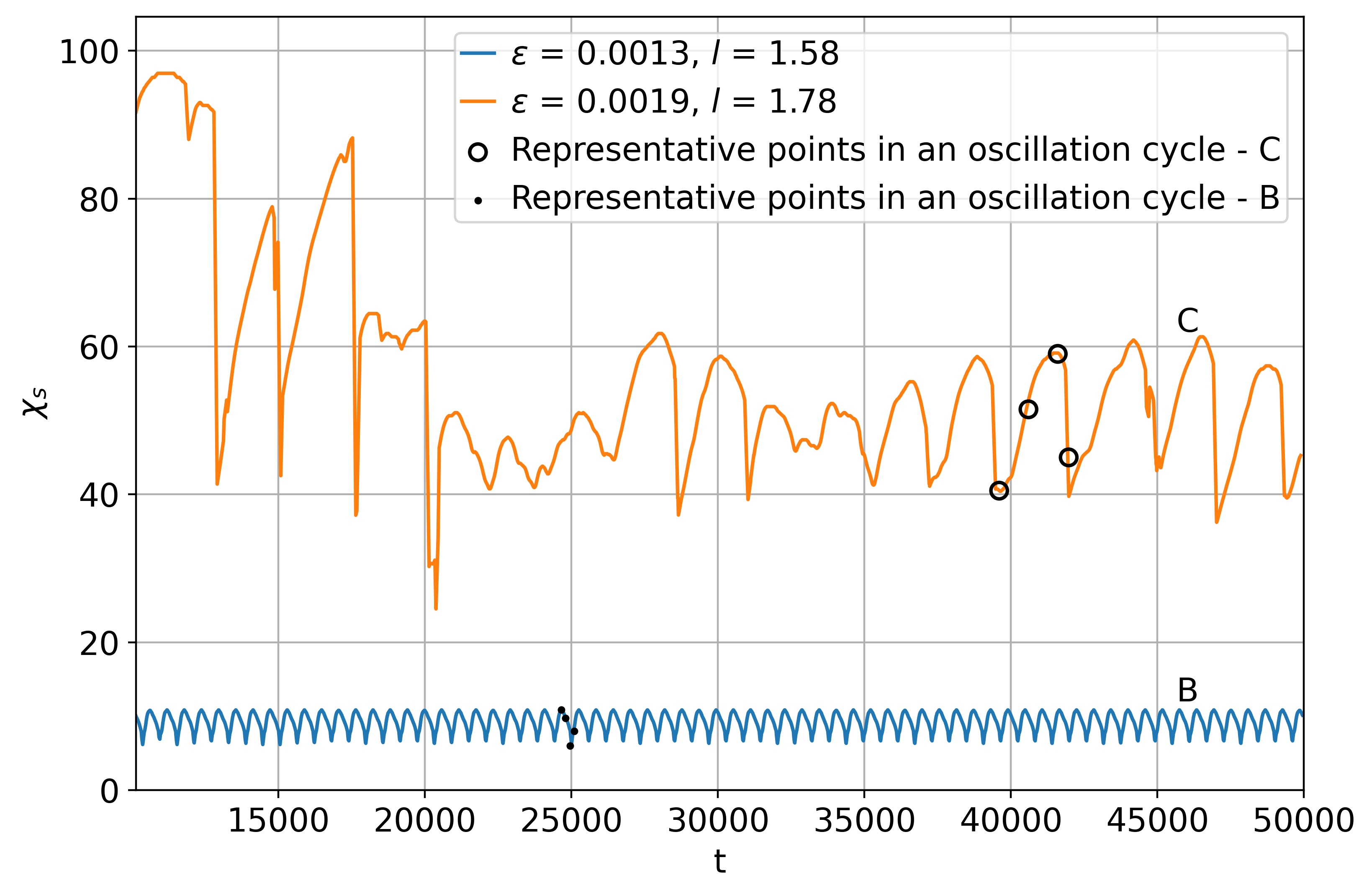}
\caption{Time variation of shock location at the equator for two cases,
one from region marked \textbf{B} and another from \textbf{C}.
Flow parameters are marked in the legend. Solutions exhibit 
steady oscillation for both the cases.
Four points in a single oscillation are marked on each plot.
We show the solution profiles at these four points in subsequent Figures.}
\label{fig:overall_oscillating_shock}
\end{figure}

\begin{figure*}
\centering
\includegraphics[width=\textwidth]{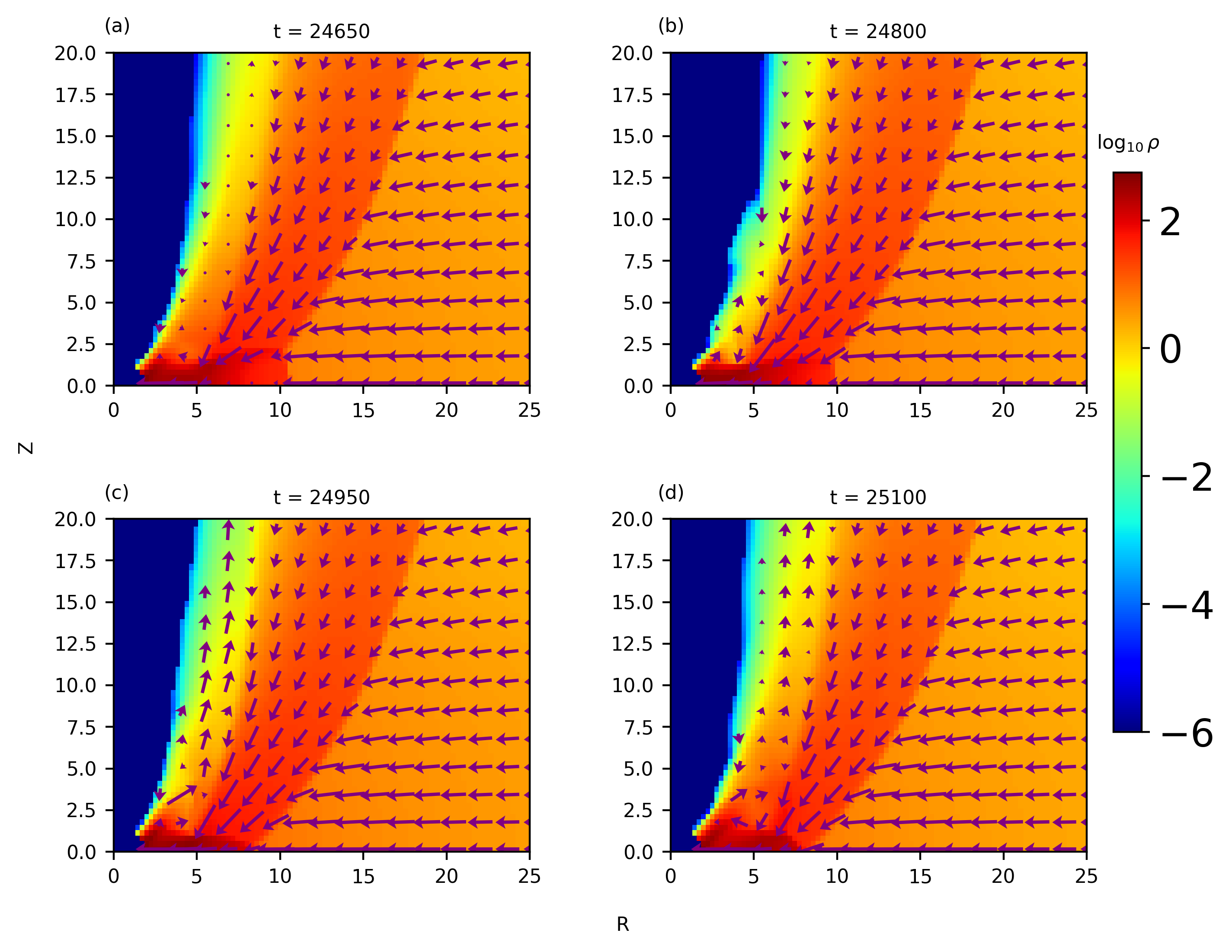}
\caption{log$_{10}\rho$ (color) - velocity vector snapshots at 
$t =$ 24650 (a), 24800 (b) 24950 (c) and 25100 (d) revealing
the accretion disk dynamics over a full
oscillation for a set of parameters from region \textbf{B}. 
These timestamps are marked by four solid black points in Fig.
\ref{fig:overall_oscillating_shock}. Simulation parameters 
$\epsilon$ = 0.0013 and $l$= 1.58 are used for this simulation. See
text for details.}
\label{fig:oscillating_shock_B}
\end{figure*}

Fig. \ref{fig:overall_oscillating_shock} shows the time variation
of shock location on the equatorial plane for two different cases,
one from region marked \textbf{B} and another from \textbf{C}.
The representative case from \textbf{B} has $\epsilon$ = 0.0013
and $l$= 1.58.
This simulation has been conducted in a simulation domain
[0:99.72]$\times$[0:49.765] using [256$\times$128] ratioed grids.
All points from this region show a similar type of
higher frequency, regular oscillations. Upon careful
observation, we notice that the shock oscillations are mostly
confined within a region close to the equator. The upper part of the CENBOL
doesn't participate in the oscillation. To demonstrate this, we
plot the density distribution at four successive time points,
indicated by four black solid dots within a single oscillation. 
The four snapshots are shown in Fig. \ref{fig:oscillating_shock_B}
(a), (b), (c), (d), and they are drawn at instances $t =$ 24650, 
24800, 24950, 25100, respectively. Color shows log$_{10}\rho$ 
distribution and arrows show velocity vectors. 
(a) shows that a parabolic shock surface touches the equator at
$R=10.8$. At subsequent times (i.e, (b) and (c)), the lower part
of the shock surface moves inward ($R=9.7$ and $R=5.8$ respectively),
and then in (d), it again starts moving outward ($R=8.02$).
During this full oscillation, the upper part doesn't move at all. 

\begin{figure*}
\centering
\includegraphics[width=\textwidth]{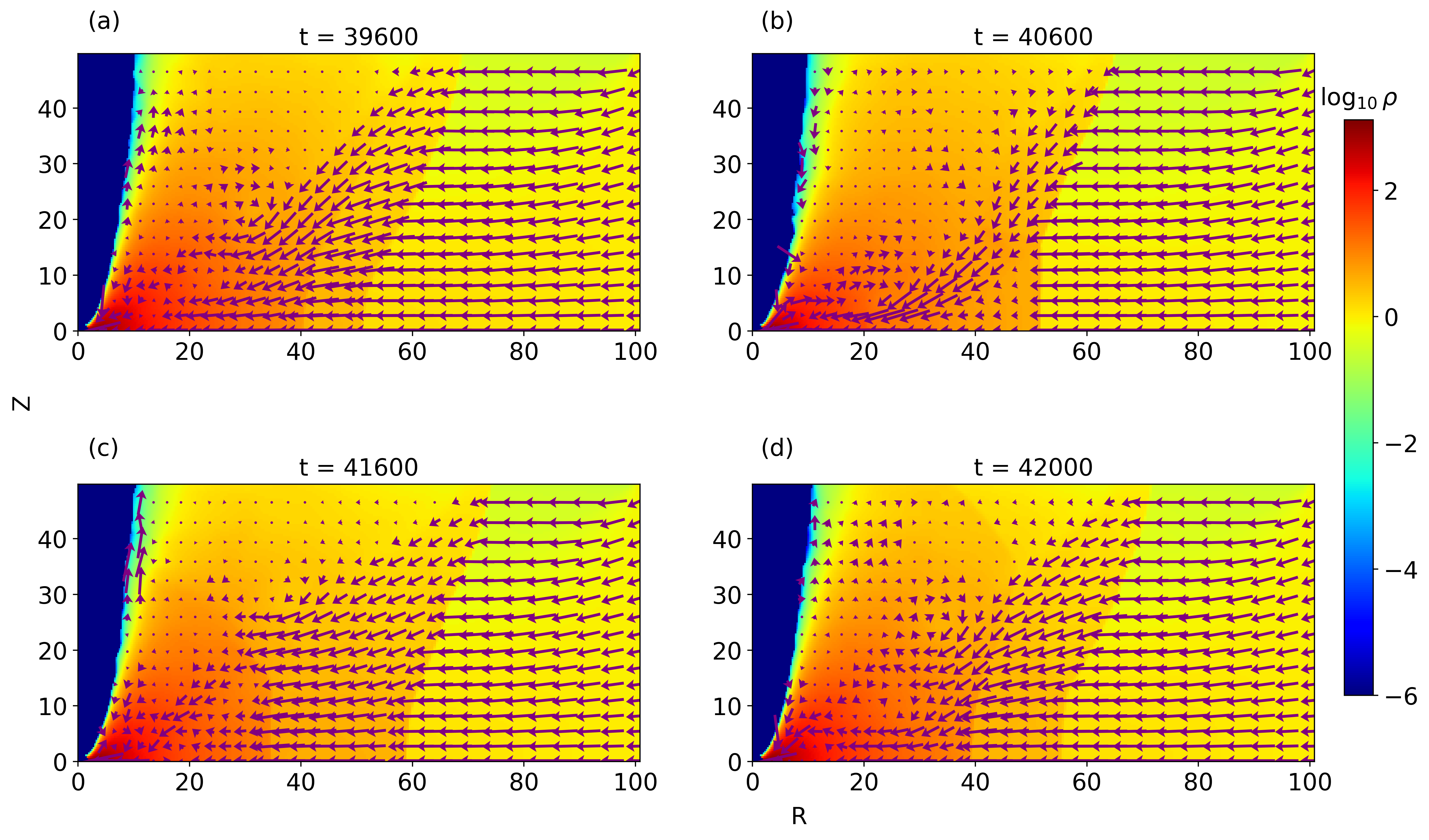}
\caption{log$_{10}\rho$ (color) - velocity vector snapshots at
$t =$ 39600 (a), 40600 (b) 41600 (c) and 42000 (d) revealing
the accretion disk dynamics over a full
oscillation for a set of parameters from region \textbf{C}. These 
timestamps are marked by four circles in Fig.
\ref{fig:overall_oscillating_shock}. Simulation parameters 
$\epsilon$ = 0.0019 and $l$ = 1.78 are used for this simulation. See
text for details.}
\label{fig:oscillating_shock_C}
\end{figure*}

In contrast to \textbf{B}, solution for points from region \textbf{C}
shows large amplitude shock oscillation as indicated
by the orange line in Fig. \ref{fig:overall_oscillating_shock}. 
For this particular case, $\epsilon$ = 0.0019 and $l$ = 1.78.
This simulation has been conducted in a simulation domain
[0:99.72]$\times$[0:49.765] using [256$\times$128] ratioed grids.
The solution shows oscillation of the entire shock surface.
We show the density distribution at four successive time points,
indicated by four black circles within a single oscillation.
The four snapshots are shown in Fig. \ref{fig:oscillating_shock_C}
(a), (b), (c), (d), and they are drawn at instances $t =$ 39600,
40600, 41600, 42000, respectively. Here, again, color shows
log$_{10}\rho$ distribution and arrows show velocity vectors.
(a) shows the shock surface touches the equator around $R=40$.
We also observe the presence of a centrifugal barrier resulting in
a sudden density increase around $r=50$. However, this is not
a shock as the flow is still supersonic.
The shock location subsequently moves outward to nearly $R=60$
in (c). At the end of the oscillation, it comes back to
around $R=40$ in (d). As one can see, a large portion of the
disk participates in the shock oscillations. We observe
the presence of large turbulent vortices in the solution for
points from region \textbf{C}. 

\subsubsection{Region \textbf{D}}
\begin{figure*}
\centering
\includegraphics[width=\textwidth]{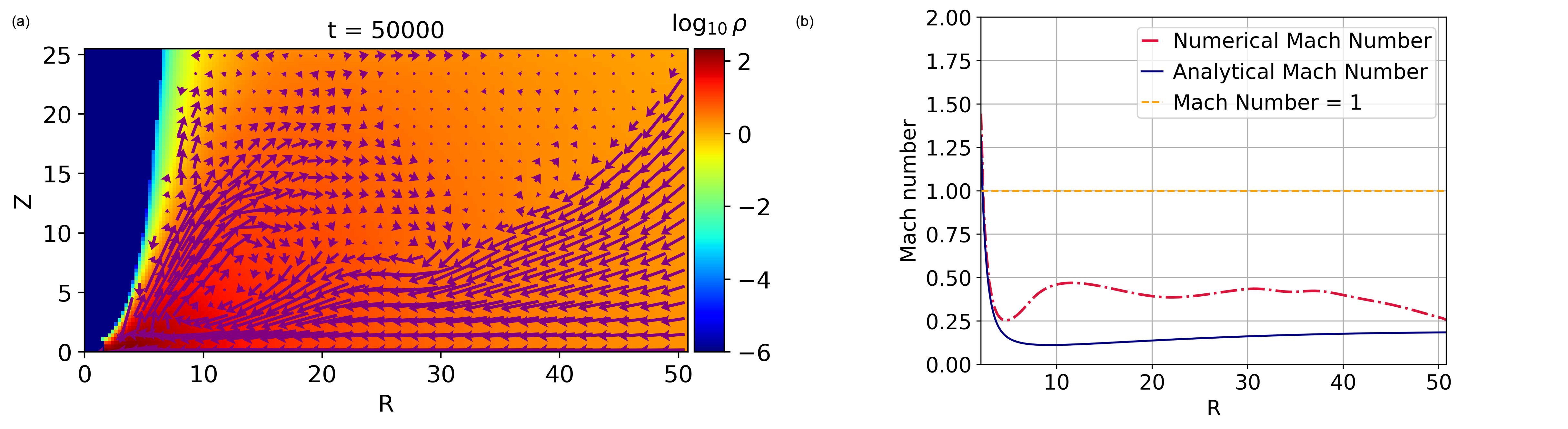}
\caption{(a) log$_{10}\rho$ overplotted with velocity vectors
for a typical point of \textbf{D} region. This is an example of
outflow-dominated, nearly non-accreting solution. (b) Radial variation
of Mach number for the corresponding case. Parameters $\epsilon$ = 0.0043 
and $l$ = 1.8 are used for this simulation. Analytically, the solution
branches passing through the outer sonic point does not reach the 
black hole horizon for this set of parameters. Only solution branches passing
through the inner sonic point reach the horizon.
Numerical simulation also demonstrates that 
the accreting matter passes through the same inner sonic point.}
\label{fig:wind_solution}
\end{figure*}

The area to the right of the shock-forming boundary, marked by
\textbf{D} in Fig. \ref{fig:overall_picture}(a), represents
outflow-dominated, nearly non-accreting solutions. A representative 
example of solution is shown in Fig. \ref{fig:wind_solution}.
$\epsilon$ = 0.0043 and $l$ = 1.8 are chosen for this one.
This simulation has been conducted in a simulation domain
[0:50.78]$\times$[0:25.308] using [148$\times$74] ratioed grids.
In Fig. \ref{fig:wind_solution}(a), we show log$_{10}\rho$ 
distribution in color, and arrows show velocity vectors. The 
snapshot is drawn at time $t =$ 50000.
The radial variation of Mach number along the equator for this
case is shown in \ref{fig:wind_solution}(b). This plot shows that 
the flow is sub-sonic at all radii except at the injection point 
and becomes supersonic only before getting accreted. The successive 
snapshots shows that these flows develop a outward propagating shock
which reaches the outer boundary of the domain, leaving behind the
sub-sonic flow. A significant amount of injected
matter can be seen to leave as outflow and form a
vortex-like structure, causing nearly no accretion of matter
onto the black hole. For higher $l$ flows, the centrifugal
barrier is sufficiently strong to stop matter from getting
accreted. 
Prior numerical simulations
in \citet{kgcb2019} also reported a similar flow profile for
higher $l$ (see Fig. 7 of this paper).

\subsection{Parameter dependence of shock location}

\begin{figure}
\centering
\includegraphics[width=\columnwidth]{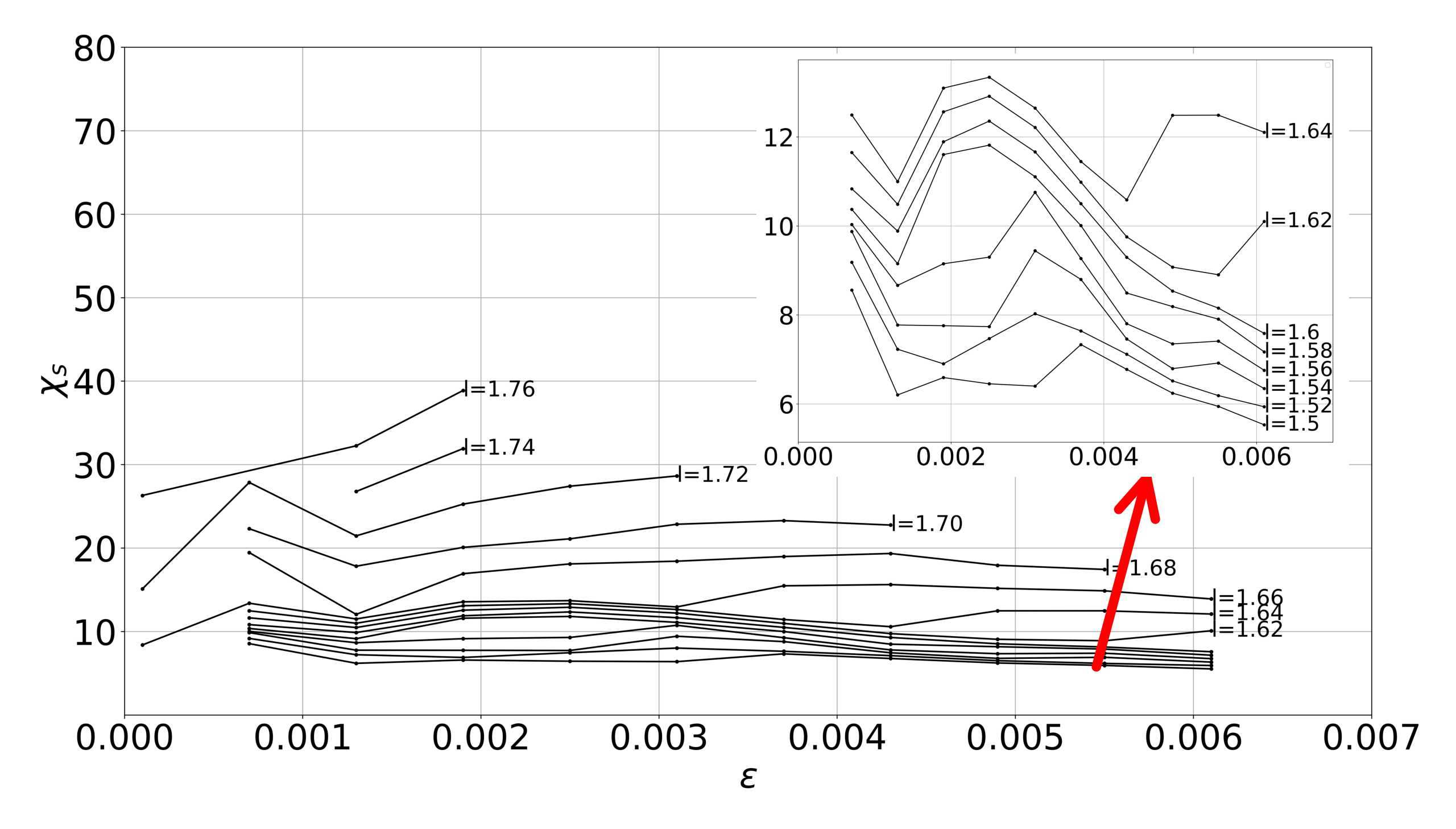}
\caption{Variation of time-averaged shock location at the equator
with $\epsilon$ for fixed $l$ values, marked on each curve.
The inset shows a zoomed-in part of the overlapping region.
We find that the shock location generally increases with $l$
for a given $\epsilon$. However, for a fixed value of $l$, 
the variation of the shock location w.r.t. $\epsilon$ is not monotonic.}
\label{fig:shock_location_variation}
\end{figure}

The location and behavior of the shock surface vary with the
flow parameters. 
We have extracted the parameter dependence of the shock location
from the simulated data. In Fig. \ref{fig:shock_location_variation},
we plot the shock-location $\chi_s$ along the vertical axis and
$\epsilon$ along the horizontal axis. Each line on this plot
shows the variation of shock location for a fixed $l$ value
(marked on the plot). 
The inset shows a zoomed-in part of overlapping region.
Note that this plot is done using only the points from region
\textbf{A} of Fig. \ref{fig:overall_picture}(a) since we find
the presence of a shock for these parameters. Also note that in many
cases, the shock surface is time-dependent. Therefore,
we show the time-averaged value of the shock location at the equator
in this plot. This plot demonstrates that for a given value of
$\epsilon$, the shock location generally increases with $l$.
For a fixed value of $l$, the variation of the shock location w.r.t.
$\epsilon$ is not monotonic though. For some part, we observe
the shock location increases with $\epsilon$, but we also observe
that the shock location either does not change much or slightly decreases
with increasing $\epsilon$ (especially for lower $l$ values).

\subsection{Parameter dependence of outflow}

In almost all our simulations, we observe the formation of outflow.
The outflow originates due to the combined effects of centrifugal
barrier and thermal pressure gradient force.
In appendix Section \ref{sec:force}, we perform a
force analysis to demonstrate this.
To quantify the outflow, we compute the outgoing mass flux $\dot{M}_{out}$
through the upper $Z$ boundary. We also compute the mass
absorption rate $\dot{M}_{abs}$ by the black hole sitting
inside $r=r_{BH}$. We generally observe that for points in
region \textbf{A} and \textbf{E}, the solutions are
absorption dominated, i.e., most of the injected matter
inside the simulation domain is accreted by the black hole.
However, as we move more towards the higher $l$ region (i.e.,
region \textbf{D}), the solution becomes outflow dominated.

\begin{figure*}
\centering
\includegraphics[width=0.32\textwidth]{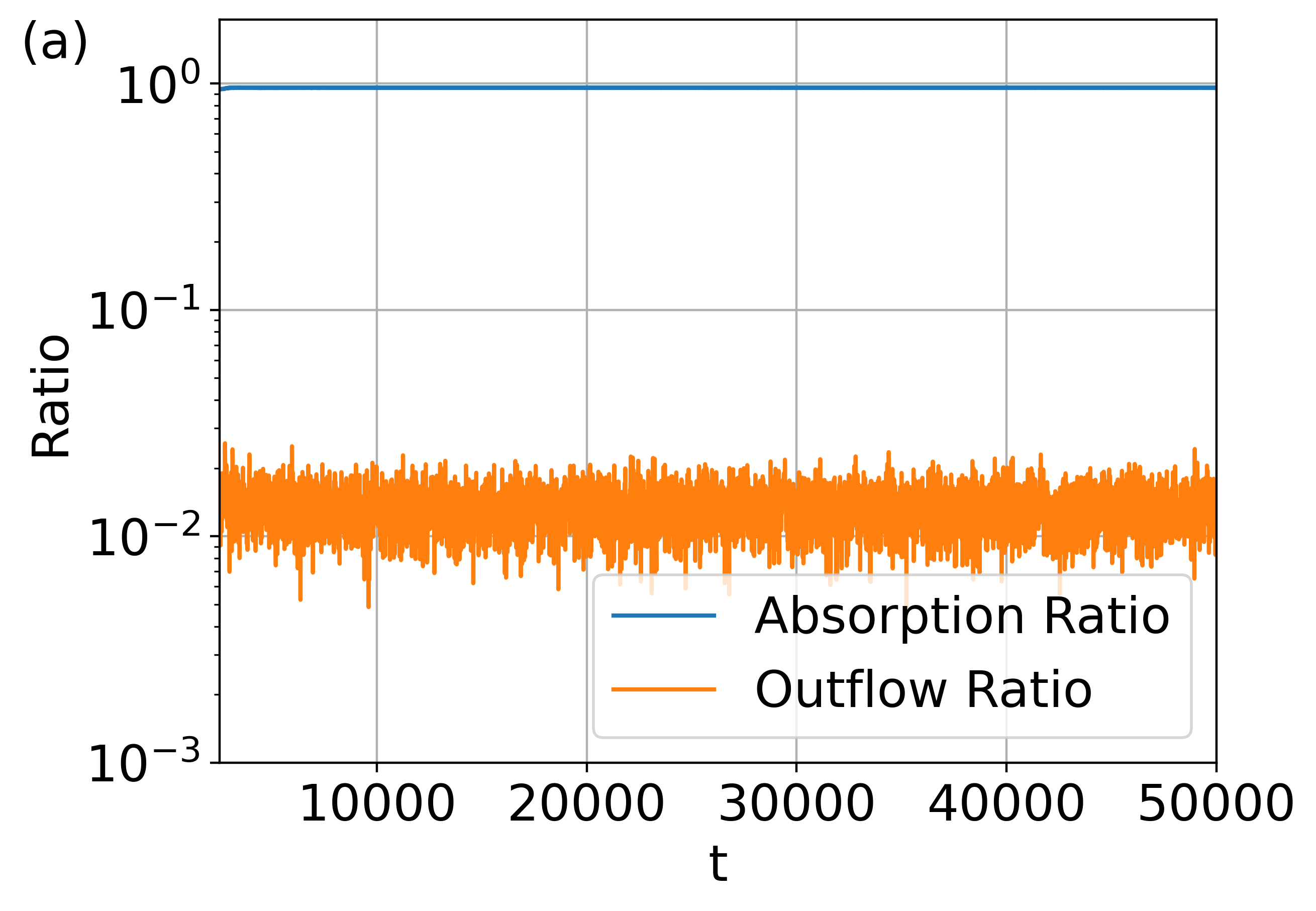}
\includegraphics[width=0.32\textwidth]{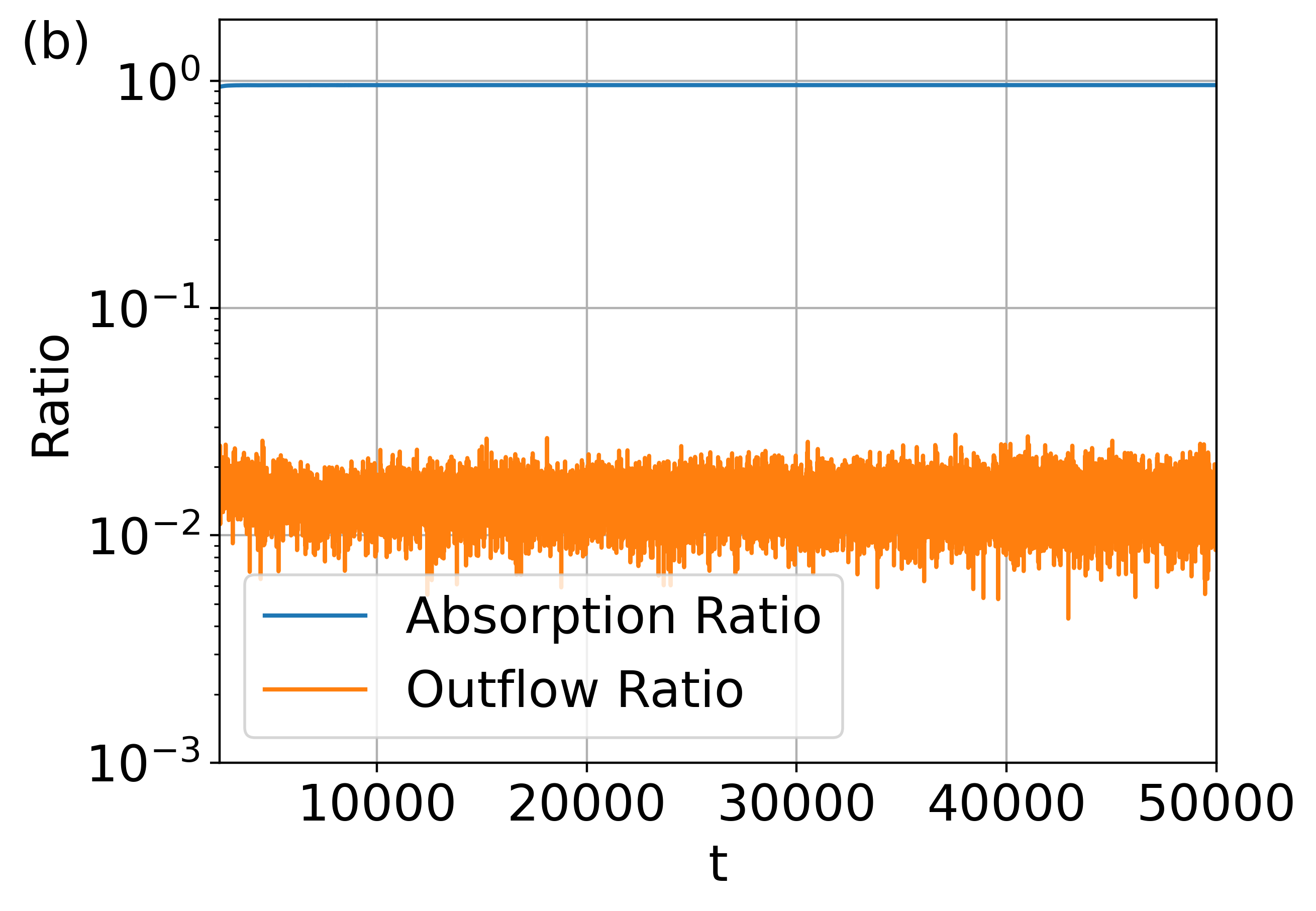}
\includegraphics[width=0.32\textwidth]{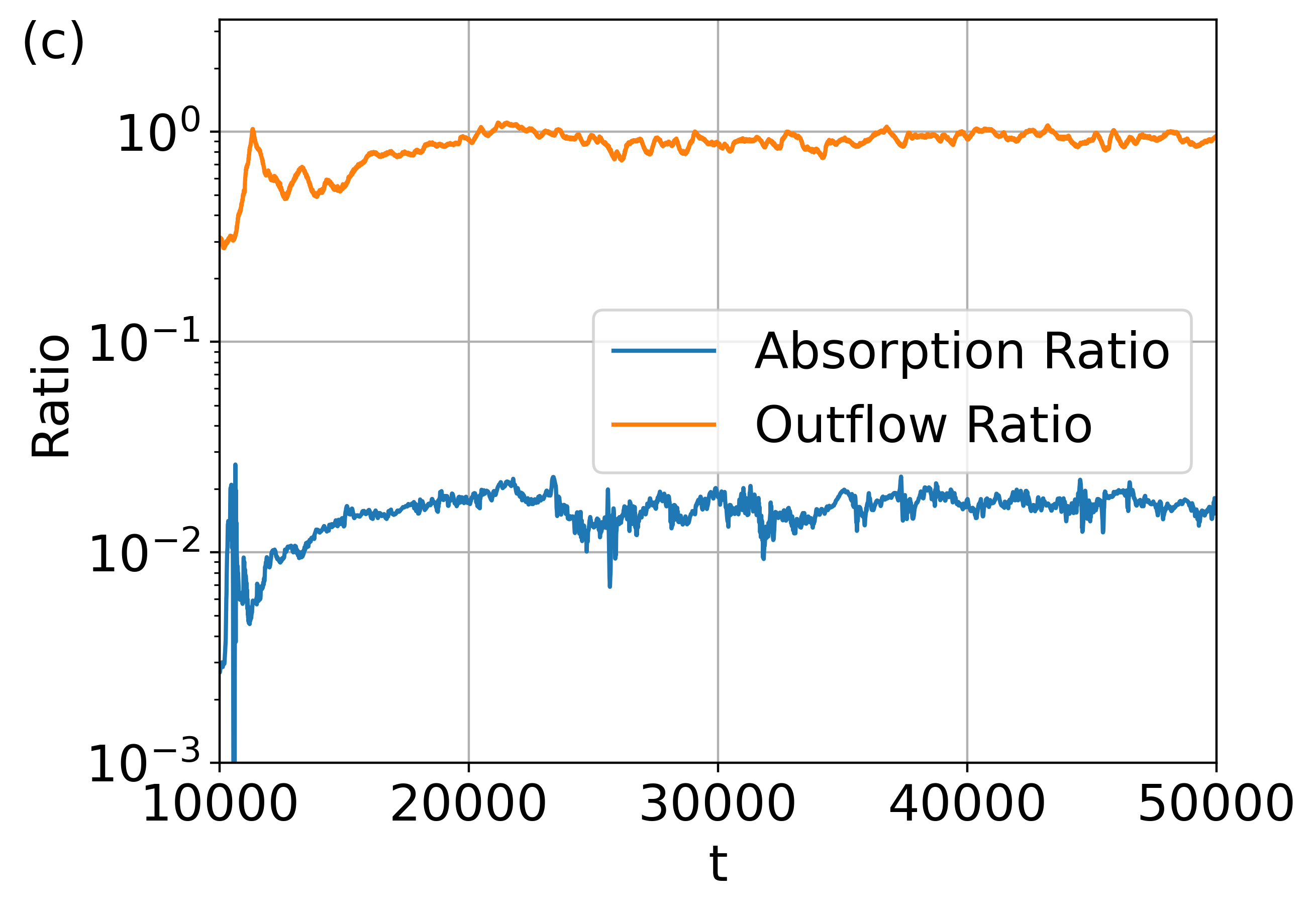}
\caption{Time variation of mass outflow rate (orange line)
and mass absorption rate (blue line), normalized by the mass 
injection rate into the simulation domain, for three
different points in the parameter space: (a) for a point
in region \textbf{A}, (b) for \textbf{E} and (c) for \textbf{D}.}
\label{fig:mass_ratio}
\end{figure*}

In Fig. \ref{fig:mass_ratio}, we show three examples of outflow and
absorption rates. The plots show time variation
of the ratio $\dot{M}_{out}/\dot{M}_{in}$ (orange line) and
$\dot{M}_{abs}/\dot{M}_{in}$ (blue line), where, $\dot{M}_{in}$ is
the inflowing mass flux through the outer $R$ boundary.
Fig. \ref{fig:mass_ratio}(a) shows these ratios for the
point from region \textbf{A}. We find that on average, nearly
95.45\% of the injected matter is getting accreted in this case.
Fig. \ref{fig:mass_ratio}(b) shows the same as (a) for the
point from region \textbf{E}. For this case, we also find that nearly
95.66\% of matter is getting accreted.
Fig. \ref{fig:mass_ratio}(c) shows the ratios for the
point from region \textbf{D}. For this case, the solution shows
the formation of large vortices, which are primarily caused by the
interaction between the incoming matter and the bounced-back matter
from the centrifugal barrier. Because of this strong centrifugal
barrier, matter cannot get accreted onto the black hole and mostly
leaves the system as outflow. We observe that more than 86\% matter
leaves the domain as outflow.

\section{Summary and Discussions}
\label{sec:5}

In this paper, we numerically simulate the non-dissipative, 
sub-Keplerian accretion flow onto a non-rotating black hole for a range of
flow parameters spanned by specific energy $\epsilon$ and specific
angular momentum $l$. Such accretion solutions are known to
harbor shock and produce a boundary layer (CENBOL) around a
black hole for certain ranges of these parameters. 
The theoretical method for shock parameter
space identification is based on whether a solution contains
a standing shock or not \citep{Chakrabarti1989a, Molteni1994a, das2001, 
mondal2006, mondal2020}.
However, there are several numerical simulations that
demonstrate that in multi-dimensional simulations, shocks can 
be standing or oscillatory and it can form for flow parameters
even outside these theoretical limits.
However, a full parameter space investigation based on purely
multi-dimensional numerical simulations are still missing.
Our current paper fills that gap.

We scan the parameter space spanned by $l$ and $\epsilon$
in the domain [1.4:1.9]x[0.0001:0.0061] using judiciously chosen
286 points. Based on the simulation results, we identify
extended ranges of these parameters that allow shock formation
(Fig. \ref{fig:overall_picture}(a)). Within this shock-forming
region, not all points show a standing shock. There are
regions that show oscillating shocks. In our analysis, we
categorized a solution as oscillatory only if the oscillation
amplitude is larger than 2$r_g$. We broadly identify
two types of oscillations. In one type (region marked \textbf{B}),
only the lower part
(closer to the equator) of the shock surface participates
in the oscillation, the oscillation frequency is higher, and
the amplitude is usually smaller.
In the other type (region marked \textbf{C}), the entire
shock surface moves, the oscillation frequency is lower, and the amplitude is usually larger. Oscillations identified in
region \textbf{B} were indicated in the theoretical analysis
of \citet{das2001}.
Additionally, similar types of oscillations as in \textbf{C}
are also reported in earlier simulations \citep{Giri2010a}.
These oscillating shocks have great implications in explaining
timing features such as QPOs.

From our simulation, we also extract the global parameter dependence
of shock location and outflow properties. We generally find
that the shock forms farther with increasing $l$ for a given 
$\epsilon$. However, variation of shock location with $\epsilon$
for a given $l$ is found to show weak dependence. Most of our
simulations naturally produce outflow due to the combined effects
of centrifugal and thermal pressure gradient forces.
The outflow rate depends strongly on $l$. All these findings
generally match the previous theoretical analyses
or the simulations.

Formation of a hotter, optically slim (optical depth $\sim$ 1) disk
as a result of shock in the sub-Keplerian flow,
naturally explains the origin of corona around a black hole. This
corona can inverse-Comptonize low energy seed photons, originating 
from a truncated Shakura-Sunyaev type thin disk \citep{Shakura1973a},
and produce a high energy tail observed in spectrally hard states.
In case when thin disk is absent, this corona can produce high energy
seed photons and self-Comptonize them to give rise to hard spectral
states.
This corona also produces outflow self-consistently which is
frequently observed in a spectrally hard state.
Our analysis demonstrates that a large parameter space actually exhibits
such corona formation. Hence, this process may be very common in
a X-ray binary system or an active galactic nuclei observed in
spectrally hard state. Our simulations also reveal that certain
region of parameters exhibit dynamic, oscillatory corona. Such
corona will induce the oscillation in the outgoing radiation
and may explain observed quasi-periodic oscillations in the high
energy band.

In passing, we note a limitation of our main result.
In this work, we simulate flow profile only in the upper half
of the $R-Z$ domain and use reflection symmetry along the equator.
Prior experience shows that unless flow has sufficiently large $l$,
flow profiles in both haves are similar. However, for a large 
value of $l$, the CENBOL may show vertical
oscillations due to strong post-shock turbulence \citep{deb2016,garain2023}
in addition to radial oscillations. In our case, large $l$ 
simulations are anyway mostly classified as oscillating or 
outflow-dominated solutions. Thus, we believe
that the parameter space boundary for shock won't be affected too much
because of our use of reflection symmetry along the equator.


The exercise presented here is done for non-rotating black holes
using pseudo-Newtonian potential. Similar parameter space
classification around rotating black holes using fully general
relativistic calculations
is also done in a few literatures \citep{Chakrabarti1996b,
Chakrabarti1996c}. As shown in these references, for corotating
flows, one can find shocks very close to a black hole. 
If such shocks oscillate
with high enough frequency, they may be able to explain
high-frequency QPOs found in a few
sources. We shall perform parameter space investigation
around rotating black holes in the future using multi-dimensional
general relativistic hydrodynamics
simulations \citep{garain2025} and the results will be reported elsewhere.
Similar parameter space investigation
can also be done using pseudo-Newtonian potential representation
for Kerr black holes
as well \citep{ck1992,bani2002,cm2006,dihingia2018} just by replacing
the potential $\Phi(r)$ in this work. However, all these potentials
have their own limitations as discussed in these references and one
has to be careful about those. Full general relativistic calculations
are free from these limitations.

Finally, we recognize that our equations and hence the results
are valid for ideal flow. More realistic flow may have various 
dissipations such as viscous
transport or radiative loss. Also, magnetic field, being very pervasive,
is expected to be present in the accretion flow. Inclusion of these
physical components would affect the flow solutions. Hence, the parameter
space may change in presence of these. However, in absence of
reliable quantitative measurement of these dissipative processes,
we generally rely on parametric representation of these affects.
Thus, the classification of accretion solutions should also depend
on the choice of these new parameters.

\section*{Acknowledgements}

We acknowledge the usage of the Kepler Computing facility,
maintained by the Department of Physical Sciences, IISER Kolkata.
We also acknowledge the usage of IUCAA's Pegasus Computing facility
for conducting a few simulations. SKG also acknowledges partial
financial support from Anusandhan National Research Foundation through
grant no. ANRF/ARGM/2025/002001/TS.

\section*{Data Availability}

The simulation data will be available from the corresponding author
on reasonable request. 



\bibliographystyle{mnras}
\bibliography{references} 




\appendix

\section{Force analysis}
\label{sec:force}

\begin{figure*}
\centering
\includegraphics[width=\textwidth]{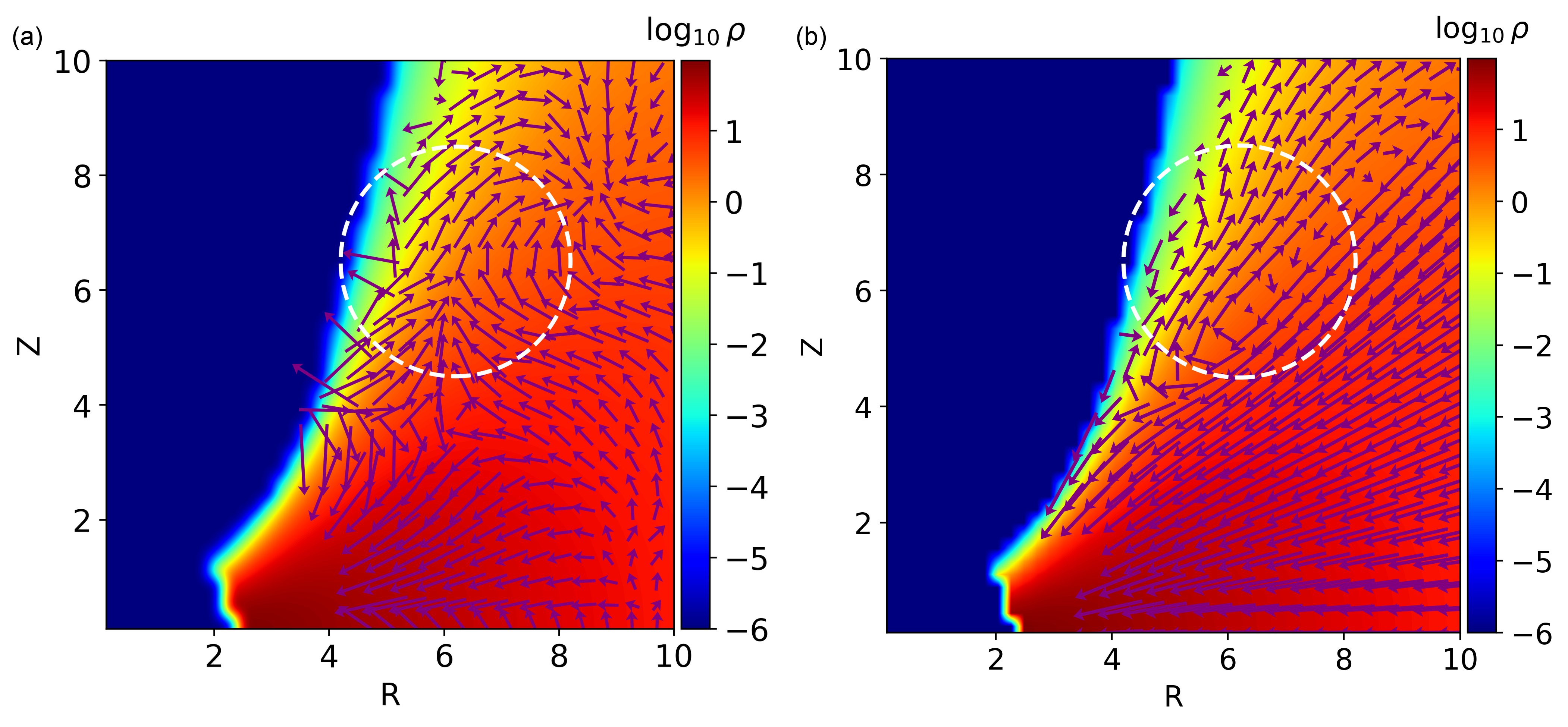}
\caption{show (a) net force vectors and (b) velocity field plotted over
log$_{10}\rho$ distribution for a point from the shock region. 
The region bounded by the white circle in (a) shows the force 
field away from the black hole. The corresponding velocity field in (b)
shows that matter is getting accelerated outward from this region.}
\label{fig:force_distribution}
\end{figure*}

In this section, we analyze the net force distribution to support 
the claim that the outflows in our simulations are generated from 
centrifugal and pressure gradient forces. For this analysis, we
choose a case with flow parameters $\epsilon = 0.0061$ and $l = 1.66$.
For this set of parameters, the flow profile is smooth with a
clear standing shock and nearly steady outflow from the post-shock
region. The two components of the net force, $F_R$ and $F_Z$ 
(see Eq. \ref{eq1}), are as follows:
\begin{align}
    F_R &= \frac{v_{\phi}^2}{R} - \frac{1}{\rho}\frac{\partial P}{\partial R} - \frac{\partial \Phi}{\partial r} \frac{R}{r}, \\
    F_Z &= -\frac{1}{\rho}\frac{\partial P}{\partial Z} - \frac{\partial \Phi}{\partial r} \frac{Z}{r}.
\end{align}
These force components are computed at each grid point.
Fig.\ref{fig:force_distribution} presents a comparative picture 
of the (a) net force field and the corresponding (b) velocity field, both 
overplotted with the log$_{10}\rho$ distribution. The length of
an arrow is proportional to the logarithmic magnitude of the vector in both plots.
We show a zoomed-in region [0:10]x[0:10] in these plots. The region
bounded by the white circle in Fig. \ref{fig:force_distribution}(a)
shows the force field away from the black hole. The corresponding
velocity field shows that matter is getting accelerated outward
from this region. Since this case shows mostly a steady solution,
the velocity vectors towards upper part of the white circle 
are already accelerated to nearly
a steady velocity by this steady force field. 
%


\bsp	
\label{lastpage}
\end{document}